\documentclass[runningheads]{llncs}

\usepackage{amsmath,amssymb,amsfonts}
\usepackage{algorithmic}
\usepackage{textcomp}
\usepackage[export]{adjustbox}
\usepackage{listings}
\usepackage{tabularx}
\usepackage{booktabs}
\usepackage{wrapfig}
\usepackage{sansmath}
\usepackage{microtype}

\usepackage[dvipsnames]{xcolor}
\usepackage{subcaption}

\definecolor{codegreen}{rgb}{0,0.6,0}
\definecolor{codegray}{rgb}{0.5,0.5,0.5}
\definecolor{codepurple}{rgb}{0.58,0,0.82}
\definecolor{backcolour}{rgb}{0.95,0.95,0.92}

\lstdefinestyle{mystyle}{
    backgroundcolor=\color{backcolour},   
    commentstyle=\color{codegreen},
    keywordstyle=\color{magenta},
    numberstyle=\tiny\color{codegray},
    stringstyle=\color{codepurple},
    basicstyle=\ttfamily\footnotesize,
    breakatwhitespace=false,         
    breaklines=true,                 
    captionpos=b,                    
    keepspaces=true,                 
    numbers=left,                    
    numbersep=5pt,                  
    showspaces=false,                
    showstringspaces=false,
    showtabs=false,                  
    tabsize=2
}

\makeatletter

\newif\ifpredicate@prolog@
\newif\ifwithinparens@prolog@

\lst@SaveOutputDef{`_}\underscore@prolog

\newcount\currentchar@prolog

\newcommand\@testChar@prolog%
{%
  \ifnum\lst@mode=\lst@Pmode%
    \detectTypeAndHighlight@prolog%
  \else
    \ifwithinparens@prolog@%
      \detectTypeAndHighlight@prolog%
    \fi
  \fi
  \global\predicate@prolog@false%
}

\newcommand\detectTypeAndHighlight@prolog
{%
  \def\lst@thestyle{\PrologAtomStyle}%
  \ifpredicate@prolog@%
    \def\lst@thestyle{\PrologPredicateStyle}%
  \else
    \expandafter\splitfirstchar@prolog\expandafter{\the\lst@token}%
    \expandafter\ifx\@testChar@prolog\underscore@prolog%
      \ifnum\lst@length=1%
        \let\lst@thestyle\PrologAnonymVarStyle%
      \else
        \let\lst@thestyle\PrologVarStyle%
      \fi
    \else
      \currentchar@prolog=65
      \loop
        \expandafter\ifnum\expandafter`\@testChar@prolog=\currentchar@prolog%
          \let\lst@thestyle\PrologVarStyle%
          \let\iterate\relax
        \fi
        \advance \currentchar@prolog by 1
        \unless\ifnum\currentchar@prolog>90
      \repeat
    \fi
  \fi
}
\newcommand\splitfirstchar@prolog{}
\def\splitfirstchar@prolog#1{\@splitfirstchar@prolog#1\relax}
\newcommand\@splitfirstchar@prolog{}
\def\@splitfirstchar@prolog#1#2\relax{\def\@testChar@prolog{#1}}

\def\beginlstdelim#1#2%
{%
  \def\endlstdelim{\PrologOtherStyle #2\egroup}%
  {\PrologOtherStyle #1}%
  \global\predicate@prolog@false%
  \withinparens@prolog@true%
}

\newcommand\lang@prolog{Prolog-pretty}
\expandafter\lst@NormedDef\expandafter\normlang@prolog%
  \expandafter{\lang@prolog}

\expandafter\expandafter\expandafter\lstdefinelanguage\expandafter%
{\lang@prolog}
{%
  language            = Prolog,
  keywords            = {},      
  showstringspaces    = false,
  alsoletter          = (,
  moredelim           = **[is][\beginlstdelim{(}{)}]{(}{)},
  MoreSelectCharTable =
    \lst@DefSaveDef{`(}\opparen@prolog{\global\predicate@prolog@true\opparen@prolog},
}

\newcommand\@ddedToOutput@prolog\relax
\lst@AddToHook{Output}{\@ddedToOutput@prolog}

\lst@AddToHook{PreInit}
{%
  \ifx\lst@language\normlang@prolog%
    \let\@ddedToOutput@prolog\@testChar@prolog%
  \fi
}

\lst@AddToHook{DeInit}{\renewcommand\@ddedToOutput@prolog{}}

\makeatother
\definecolor{blue3F6696}{HTML}{3F6696}
\definecolor{blueADC1D7}{HTML}{ADC1D7}
\definecolor{blueD3E0EE}{HTML}{D3E0EE}
\definecolor{darkyellow}{rgb}{0.75,0.75,0.0}
\definecolor{lightgray}{rgb}{0.8,0.8,0.8}
\definecolor{orange}{rgb}{1.0,0.5,0.0}
\definecolor{verylightgray}{rgb}{0.95,0.95,0.95}
\definecolor{green-verystrong}{HTML}{4F6228}
\definecolor{green-strong}{HTML}{76923C}
\definecolor{green-medium}{HTML}{C2D99B}
\definecolor{green-low}{HTML}{D6E3BC}
\definecolor{green-no}{HTML}{FFFFFF}

\newcommand\PrologPredicateStyle{}
\newcommand\PrologVarStyle{}
\newcommand\PrologAnonymVarStyle{}
\newcommand\PrologAtomStyle{}
\newcommand\PrologOtherStyle{}
\newcommand\PrologCommentStyle{}

\definecolor{PrologPredicate}{RGB}{000,031,255}
\definecolor{PrologVar}      {RGB}{024,021,125}
\definecolor{PrologAnonymVar}{RGB}{000,127,000}
\definecolor{PrologAtom}     {RGB}{186,032,032}
\definecolor{PrologComment}  {RGB}{063,128,127}
\definecolor{PrologOther}    {RGB}{000,000,000}

\renewcommand\PrologPredicateStyle{\color{PrologPredicate}}
\renewcommand\PrologVarStyle{\color{PrologVar}}
\renewcommand\PrologAnonymVarStyle{\color{PrologAnonymVar}}
\renewcommand\PrologAtomStyle{\color{PrologAtom}}
\renewcommand\PrologCommentStyle{\itshape\color{PrologComment}}
\renewcommand\PrologOtherStyle{\color{PrologOther}}

\lstdefinestyle{Prolog-pygsty}
{
  language     = Prolog-pretty,
  upquote      = true,
  basicstyle=\ttfamily\footnotesize,
  breakatwhitespace=false,   
  stringstyle  = \PrologAtomStyle,
  commentstyle = \PrologCommentStyle,
  literate     =
    {:-}{{\PrologOtherStyle :-}}2
    {,}{{\PrologOtherStyle ,}}1
    {.}{{\PrologOtherStyle .}}1
}

\begin{document}

\title{Monitoring Auditable Claims in the Cloud 
}

\author{Lev Sorokin, Ulrich Schoepp}
\institute{fortiss GmbH, Munich, Germany\\
\email{\{sorokin,schoepp\}@fortiss.org}
}
\maketitle

\begin{abstract} 
When deploying mission-critical systems in the cloud, where deviations may have severe consequences, the assurance of critical decisions becomes essential.
Typical cloud systems are operated by third parties and are built on complex software stacks consisting of e.g., Kubernetes, Istio, or Kafka, which due to their size are difficult to be verified. Nevertheless, one needs to make sure that mission-critical choices are made correctly.
We propose a flexible runtime monitoring approach that is independent of the implementation of the observed system that allows to monitor safety and data-related properties.
Our approach is based on combining distributed Datalog-based programs with tamper-proof storage based on Trillian to verify the premises of safety-critical actions. The approach can be seen as a generalization of the Certificate Transparency project.  
 We apply our approach to an industrial use case that uses a cloud infrastructure for orchestrating unmanned air vehicles.
\end{abstract}


\section{Introduction}





Cloud systems are used in various domains, e.g., in healthcare, entertainment, in the financial or the automotive sector. Following the HashiCorp cloud report\footnote{https://www.hashicorp.com/state-of-the-cloud} over 94\% of all enterprises world-wide use cloud services.
Deployment of systems in the cloud promises cost savings, scalability -- by automatic provisioning of resources depending on the request load -- and high availability.

While cloud technologies have many advantages, using them for mission-critical applications presents many challenges. We need to address several risks, such as hardware, software faults or attacks that can have an impact on the integrity of processed information or the availability of a service. However, cloud software stacks based on components like Kubernetes or Kafka are complex\footnote{\url{https://www.influxdata.com/blog/will-kubernetes-collapse-under-the-weight-of-its-complexity/}} and are updated frequently, which makes assurance tasks difficult.

If it is not feasible to verify the implementation, then one approach to guaranteeing the safety of critical decisions is to track the events and actions of the implementation in an independent runtime monitoring layer and to verify the safety of decisions using the collected data.

Runtime monitoring is already standard in cloud systems in the form of Security Information and Event Management (SIEM) \cite{miller2011harper,BachaneCloudSIEM16}. One continuously aggregates log data from all cloud system components and correlates and analyses it to identify potential security issues. However, information flows only from the components to the SIEM and this allows one to identify problems only after the fact. To make sure that mission-critical decisions are made safely, we want to perform the analysis during system operation and provide the results to the components so that they can make their decisions safely. An example instance of this idea is the  Certificate Transparency\footnote{https://certificate.transparency.dev/} project, where HTTPS certificates are stored in a public ledger that allows their verification, and browsers can make the decision on whether to accept a certificate based on information in such a ledger.

The problem of analysing the behavior with \textit{runtime monitoring} has been the topic of extensive work, see, e.g., \cite{CassarRV17}. For example, monitoring approaches exist based on generating monitors from automata or using event stream processing \cite{rtConversationChechik,cotroneo2020runtime,BratanisMonitoring10,DeckerMedMon14}. To assure the safety of mission-critical decisions, we are interested in a particular monitoring problem that does not seem to be accounted for very well by such approaches. We want to record the relevant history of data leading up to a mission-critical decision and then verify that this data supports the decision before executing it. For example, in a cloud service that optimizes the operation of a world-wide fleet of unmanned air vehicles, we want to verify all preparation steps before starting any drone. To do this, we typically need data from the event history, e.g., the JSON body of API requests, that finite state automata abstract.

In addition, documentation/auditability of process steps is becoming more and more important, as seen in regulatory requirements for the privacy protection (GDPR) or the  protection of health data (HIPAA), among others. 
Traceability of events is in the interest of a service provider to have a convincing argument in case of disputes.
Existing approaches that consider auditability, such as \cite{ahmad_secure_2019,van_hoye_logging_2019,Prybila2020RuntimeVF}, are based on distributed ledger technology and are only applicable to low-volume processes, such as logistics or manufacturing, because of high transaction latencies and low transaction rates (in the order of~1000 transactions per second). 





We present an approach, that builds on ideas of the Certificate Transparency project. 
We generalized the approach to monitor safety properties in a cloud system, since it is not directly applicable to monitor events in an arbitrary cloud system. Our approach is characterized by the following contributions: 1) We introduce \texttt{Cyberlog}, a Datalog-based language for specifying the system state and properties in a distributed system. The language allows us to specify temporal properties, parallel process instances and invariants about data in messages. 2)  A property is monitored by the deployment of distributed \textit{security monitors} which capture all messages sent between components during runtime of the cloud system and analyze these events continuously. This approach allows to generate a view on the systems state during execution continuously without having access to the application code, i.e., it is not intrusive. System components have access to the Cyberlog-claims, in particular the result of the analysis of all available information, and can use this knowledge in their decisions.
3) We integrate Trillian/Rekor into the monitoring architecture for the tamper-proof logging of claims to allow an indisputable/auditable documentation of events. 
4) We provide a revision model to make reasoning with claims feasible. A \textit{knowledge revision} is necessary because the size of the generated log data at each distributed monitor otherwise does only increase.
We demonstrate the applicability of our approach on a prototypical cloud system from the avionics domain for booking transportation services with unmanned air vehicles.


\section{Motivating Example}
\label{sec:use-case}


\begin{figure*}[t]
    \centering
    \includegraphics[width=\linewidth]{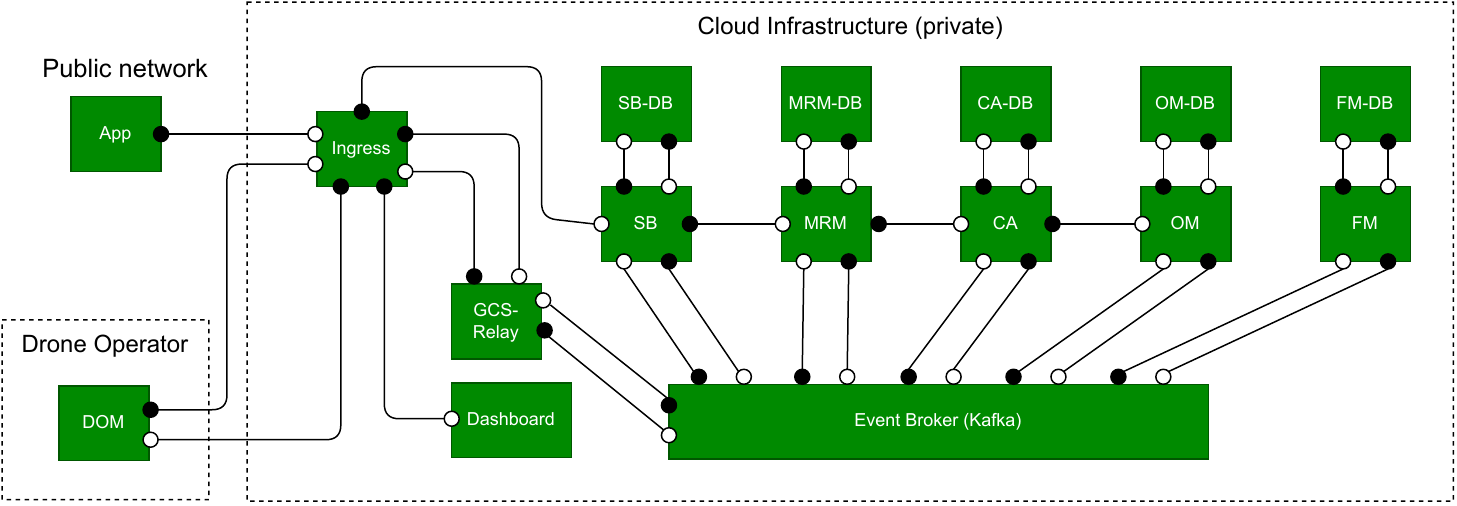}
    \caption{RTAPHM digital platform system architecture}
    \label{fig:rtaphm}
\end{figure*}

We motivate our approach with a use case from the RTAPHM project\footnote{https://www.fortiss.org/forschung/projekte/detail/rtaphm}.
Unmanned air vehicles (UAVs) have gained high popularity in the last few years, their applications range from cargo transportation, to search-and-rescue operations or environmental monitoring\footnote{https://droneii.com/237-ways-drone-applications-revolutionize-business}.
The RTAPHM project develops a cloud platform for orchestrating UAVs for such uses, in particular for the transportation of transplant organs\footnote{https://www.cbc.ca/news/canada/toronto/first-lung-transplant-drone-1.6208057}. Such applications are subject to legal auditability requirements. For example, \S486.346 of the US Code of Federal Regulations requires organ transport operators to ``develop and follow a written protocol for packaging, labeling, handling, and shipping organs in a manner that ensures their arrival without compromise to the quality of the organ.''

The platform consists of several micro-services as shown in Fig.~\ref{fig:rtaphm}: 
The Service Broker (SB) for checking the correctness and authorization of requests; the Multi-Resource-Manager (MRM) for managing available resources; the Cognitive Assistant (CA) for optimizing the allocation of resources; the Fleet Manager (FM) for predictive maintenance of UAVs; and the Operation Manager (OM) for triggering the flight preparation tasks and launches of UAVs. Each service has a local database. The services communicate both via REST APIs and also event-based using the Kafka message broker. 

The platform has an interface to Drone Operators (DOs), which are in charge of executing planned UAV missions. DOs deploy a Drone Operation Manager (DOM) component, which automates this task, using their own infrastructure. The DOM is responsible for executing UAV missions and for regularly sending status information (e.g., position, velocity) about UAVs on missions to the cloud platform. In addition, the cloud platform is connected to a public web application (App) to receive service requests from users. The connection to both the DOM and the App is realized using an ingress gateway, what is common in cloud infrastructures.
We describe in the following the typical workflow in the cloud platform for booking a UAV.

\textit{Organ transport example.} A user requests an organ transport from destination $A$ to destination $B$, the SB checks the request and forwards it to the MRM and CA to select a UAV as well as assets as transport boxes, and personnel to meet the request constraints (e.g., delivery-time, organ weight).
The user receives as a response a set of booking options together with prices. After he has selected one option the request turns into a mission, the personnel is triggered by the OM to prepare the drone and calculate waypoints for the flight. If the preparation is completed, the DO receives a ready-to-fly (RTF) request with the waypoints to launch the drone. 

We next formulate some example properties that shall not be violated when scheduling UAVs for organ transportation. 
We assume that these requirements are derived from a careful safety analysis of the system, e.g., using an STPA \cite{LevesonSTPA04}. 


\begin{enumerate}
\item \textit{Process requirements}: A DO wants to start a mission only if all required steps for a mission start have been completed and documented, e.g, a service request is documented, the booking steps have been completed by all components and the drone has been prepared correctly by ground staff. 

\item \textit{Temporal requirements}: The time between receiving and sending a message in the system should not exceed a threshold. If an RTF-message arrives delayed at the DO, as a consequence the organ will be damaged. 

\item \textit{Data-related requirements}:  The waypoints in the RTF message have to match the waypoints of one registered UAV base. Otherwise, an attacker could modify waypoints in the RTF message that are sent to DO.

\end{enumerate}

Our goal is to verify such properties independently of the implementation and to provide the results to the system component. The DOM should be able to rely on all process requirements being checked and documented before starting a UAV.
In the following sections we explain our approach to achieving this.


\section{Monitoring Approach}
\lstset{style=Prolog-pygsty}

\begin{figure}
\centering
\begin{subfigure}{.49\textwidth}
    \centering
    \includegraphics[scale=0.45]{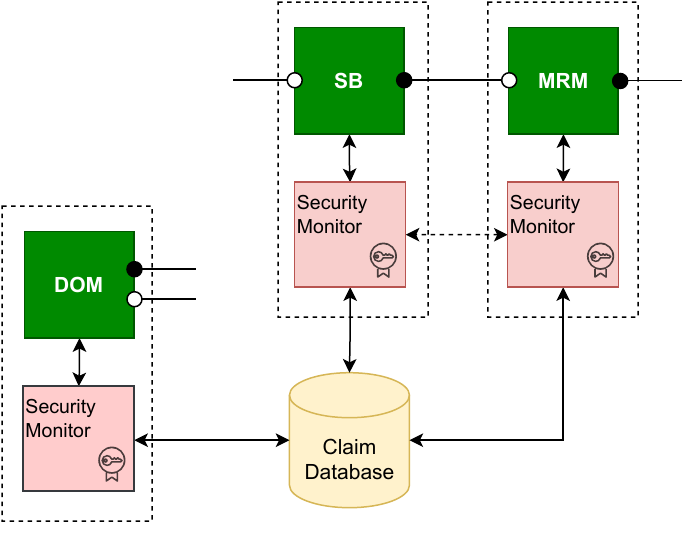}
    \caption{Monitoring with security monitors}
    \label{fig:overview-sec-monitoring}
\end{subfigure}
\begin{subfigure}{.49\textwidth}
    \centering
    \includegraphics[scale=0.5]{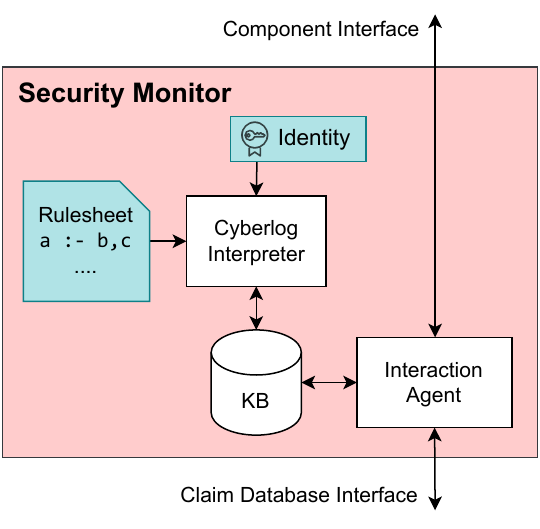}
    \caption{Security monitor architecture}
    \label{fig:monitor-architecture}
\end{subfigure}%
\caption{Overview Security Monitoring}
\label{fig:test}
\end{figure}

Our proposed monitoring system adds a verification layer to the system. Security monitors are added to the  components of the systems, as Fig.~\ref{fig:overview-sec-monitoring} shows. These monitors collect events during system operation, they analyse the data to verify requirements during operation, and they provide the results of their analysis to the components. To verify process requirements, the monitors must be able to observe the interactions between multiple components. To this end, they communicate claims about the process execution with each other. Each monitor has an assigned identity and claims are exchanged in the form of cryptographically signed attestations. For accountability and documentation of processes, the monitors communicate with a claim database that stores the exchanged claims in a tamper-proof manner.

The three main aspects of the monitoring system are a specification language for defining requirements, a security monitor and tamper-proof claim storage. In the following section we outline them in more detail.



\subsection{Specification Language} 
In order to monitor the dynamic behaviour of a system, we need to specify the required behaviour/properties in a specification language. For this, we propose an extension of Datalog \cite{datalogCeri89}, called \emph{Cyberlog}. 
Our language is based on Datalog, because Datalog allows to specify an abstract view on the overall systems state by using \textit{rules} that are applied over a dataset.
Datalog has a well-defined semantics and is often used as a query language for declarative databases, and has been already applied to specify the behaviour of distributed systems \cite{DeclarativeCloud,ChuDSN07}. 
Datalog requires an interpreter that operates on a \textit{knowledge base (KB)} and uses the predefined rules to incrementally update the KB with new reasoned facts by applying the rules on the given data in the KB. When new data, such as an API request in the web service is registered and logged in the KB, the reasoning automatically takes place. This allows us continuously to update the view on the systems state when new system events are stored in the KB.

We describe briefly the syntax of Cyberlog:
A Cyberlog \textit{rule} has the form \lstinline+ a :- b_1, ..., b_n+, where \lstinline+a, b_i+ are atoms.  The atom \lstinline+a+ is called the \textit{head} of the rule, while the \lstinline+b_i+ form the \textit{body} of the rule. In Cyberlog, the atoms are attestations of the form \lstinline+ principal attests symbol(t_1,...,t_n)+, where \lstinline+principal+ is a term denoting the identity of a component or person associated with the system, \lstinline+symbol+ is a \textit{predicate} and the \lstinline+t_i+ are \textit{terms}. A term can be a variable (starts with an uppercase character), or a constant, such as string literals \lstinline+'text'+ or numbers. 



A Cyberlog program, called \textit{rulesheet}, consists of a set of Cyberlog rules. Such a program is intended to be executed by a single principal. We therefore allow the user to write short \lstinline+symbol(...)+ for \lstinline+self attests symbol(...)+, where \lstinline+self+ is the identity of the principal executing the program. We require that the head of rules may only contain such self-attestations, so that principals cannot make claims for others.

An example rulesheet for the principal \lstinline+'SB'+ is given in Listing~\ref{lst:rulesheet} below. It begins with a declaration of the identities of components, where we show the details only for SB. Identities are represented through a X509 digital certificate (RFC 5280) and the rulesheet specifies the subject and issuer information of the certificate. 

\begin{lstlisting}[caption={Example Cyberlog program}, label=lst:rulesheet]
 'SB':  Subject: 'C=DE, ST=Hamburg, L=HH, O=ZAL, CN=SB'
        Issuer: 'C=DE, O=Lets Encrypt, CN=R3'
 'MRM': ...
 'OM': ...
 'CA': ...

 // interpretation of events
 request(RequestId, Data, Time) :-
   postRequest('/servicerequest', Time, Data),
   get_param_int(Data, 'request_id', RequestId).
    
 // workflow
 good_rtf_exists(RequestId, AircraftId) :-
   'SB' attests request(RequestId, Data, TimeRequest),
   'MRM' attests feasible_config(RequestId, AircraftId),
   'OM' attests tasks_done(RequestId, AircraftId),
   'OM' attests ready_to_fly(RequestId, AircraftId, Data, TimeRTF).
   
 // time
 delayed_rtf(RequestId, DelayTime, SentTime) :- 
   'OM' attests ready_to_fly(RequestId, AircraftId, DataRTF, TimeRTF),
   'CA' attests mission_confirmed(RequestId, Data, SentTime),
   DelayTime == TimeRTF - SentTime,
   DelayTime > 1000.

\end{lstlisting}
The rule in line 13 formalizes property~1 from Section~\ref{sec:use-case}. It specifies that an RTF message is considered as \textit{good} (modelled by \lstinline+good_rtf_exist+) if SB has registered a Service Request (\lstinline+request+), MRM has provided an appropriate booking option (\lstinline+feasible_config+), and if all necessary UAV preparation tasks (line 16) have been done by OM (\lstinline+tasks_done+). The claim \lstinline+good_rtf_exist+ is generated when the claims in the body of the rule are all available in the monitor's KB.

The rule in line 20 is an example of a time-related property. It verifies whether an RTF message  issued by the DO is delayed (threshold is 1000ms). The rule generates a \lstinline+delayed_rtf+ fact whenever a \lstinline+ready_to_fly+ fact exists, CA has confirmed the booking request, and the time between the sent RTF and confirmation exceeds the threshold.

Lines 7-10 demonstrate the definition of views on the available data using Cyberlog, which is in our example the modelling of a Service Request. 
The parameter \lstinline+RequestId+ is a unique identifier for a service request to distinguish between different requests, \lstinline+Time+ is the time when the request was made, and \lstinline+Data+ is the content of the message in the request (line 8).
The body of the rule in lines 8-10 declares that a \lstinline+postRequest+ claim must exist, which represents an API request events. Cyberlog also provides predicates for working with JSON data, which is ubiquitous in cloud systems. In line 10, where a request identifier is extracted from the request data \lstinline+Data+ (which holds JSON data) and bound to the \lstinline+RequestId+ variable.





\subsection{Security Monitor}

As Fig.~\ref{fig:monitor-architecture} shows, each monitor is configured with an identity and a rulesheet. The monitor has an integrated \textit{Cyberlog-interpreter} for the evaluation of rules and a \textit{local knowledge base} (KB), where Cyberlog facts are stored. 

The monitor continuously executes the rule from the rulesheet. Initially the KB is empty. There are three ways in which facts can be added to the KB:

\begin{itemize}
\item A fact is derived by applying a Cyberlog rule. 

\item A claim is passed directly to the monitor, e.g., a claim representing an event.

\item The monitor's interaction agent loads new facts from the claim database, a tamper-proof storage of claims. This is e.g., done when other monitors add facts to the claim database, which the monitor has not yet received. 
\end{itemize}

The monitor also provides a query interface where the services may request information from the knowledge base. For instance, in our use-case, the DOM may query \lstinline+good_rtf_exists+ before starting a drone.




\subsection{Tamper-proof Claim Database and Auditability} 
\label{sec:monitor-auditability}


The monitoring system guarantees that for each claim in the knowledge base of a security monitor, the complete chain of how it was derived can be reconstructed later.
To achieve this, all Cyberlog rules and generated facts are stored in the tamper-proof claim database which shall provide the following guarantees:
1) \textit{Immutability}: Once data has been stored in the database, it cannot be changed afterwards. 2) \textit{Protection against manipulation}: Manipulation of the database can be detected. 3) \textit{Inclusion proofs}: The database provides cryptographic means to verify that a given claim is really included in it\footnote{https://transparency.dev/verifiable-data-structures/}. The security monitors use inclusion proofs to make sure that only recorded claims are used in reasoning.
Inclusion proofs are necessary, e.g., to cope with the situation where an attacker attacks the interface of the claim database so that submitted claims are not stored or stored in a modified form, but are reported to the security monitors being stored in the original form.

Auditability of claims is then enabled by storing evidence for claims in the database: (1) For claims derived using rules, the evidence is the instance of the rule used to derive the claim. E.g., for a claim \lstinline+request(id,data,time)+ the evidence would be the rule in line 9 with the claims \lstinline+postRequest('/servicerequest', time, data)+ and \lstinline+get_param_int(data, 'request_id', id)+ in Listing \ref{lst:rulesheet}. (2) \label{item:claim log}
For claims loaded from the claim database, the evidence is an inclusion proof of the claim in the claim database. (3) Claims which have been submitted directly to the KB, such as \lstinline+postRequest(...)+ (which represents an API request), are atomic assertions that cannot be verified further. 
\section{Claim Revision Control}

We have outlined how Cyberlog allows us to track events and to automatically derive auditable claims about the execution of a distributed application. While flexible, the approach is naive with respect to the resource consumption of monitors: The KB of the individual monitors will only grow larger during the operation of the system. For the continuous operation of the security monitors, it is essential to be bound the amount of data in the local KBs. 

What we need is a mechanism for retracting claims from the local KBs that are no longer needed. 
There seems to be no general criterion to decide when a claim will not be needed anymore for reasoning, so it does not appear that claims can be removed fully automatically. Instead, we propose an approach to give the user the power to control the local KB explicitly. 
We provide the user with mechanisms to retract claims as part of the \textit{rulesheet} specifications.

Our approach can be seen as a generalization of Dedalus \cite{Dedalus}, an extension of Datalog with a notion of time. Dedalus models logic programming with a discrete notion of times. All claims are annotated with a time and there are rules such as \lstinline+next a :- b, c+ to express the transitions of time. The rule expresses that if \lstinline+b+ and \lstinline+c+ are true at some time~$t$, then \lstinline+a+ will be true at time~$t+1$. For monitoring applications, one would be interested in storing only the claims of the current time.

One of the main conceptual contributions of this paper is to generalize Dedalus to a workflow language for distributed claim revision control. We replace the time indices by revision in a revision system inspired by version control systems such as git or mercurial. 

\subsection{Revision Model} 

We propose a (knowledge) revision model to account for the change of claims over time. The idea is to manage claims much like git manages files. The new knowledge generated by the monitors in their local KBs is considered as a \textit{staging} revision of claims. This revision is completed by \textit{committing} it to the claim database. Once stored in the claim database, revisions cannot be modified. However, it is possible to revise a revision, e.g., to revoke claims, by committing a new revision that supersedes it.

\begin{figure}[htpt]
	\centering
 \begin{minipage}[t]{.45\textwidth}
\includegraphics[scale=.45,valign=t]{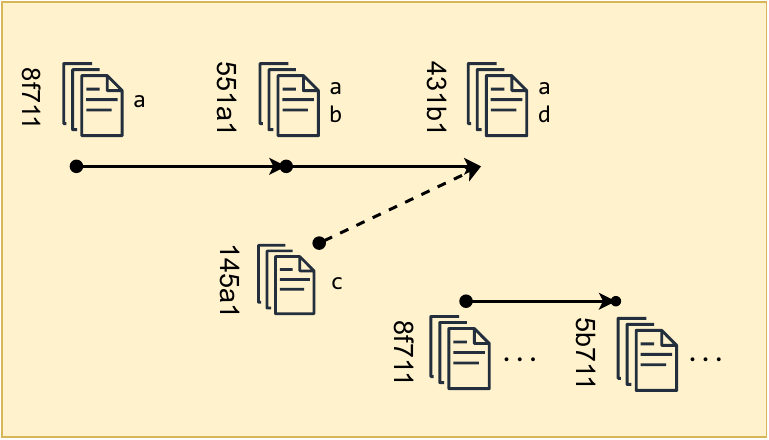}
 \end{minipage}
 \begin{minipage}[t]{.5\textwidth}
\begin{verbatim}
    Revision 431b1:
    - owner: SB
    - supersedes: 551a1
    - includes: 145a1
    - rules: <reference to rules>
    - claims: 
        - a (evidence: signed by SB)        
        - d (evidence: derived by 
              rule d :- b, c) 
\end{verbatim}
\end{minipage}
 	\caption{Revision model with superseded and included revisions of claims}
	\label{fig:revision-model}
\end{figure}

Fig.~\ref{fig:revision-model} illustrates the main concepts in the revision model. A revision (black dot) stores a set of Cyberlog claims. It is identified in the claim database by a unique ID, typically a hash like 431b1. Our revision model accounts for two possible relations between revisions: a revision can \textit{supersede} another revision (represented by a solid arrow, e.g., 551a1 supersedes 8f711) or it can \textit{include} another revision (represented by a dashed arrow, e.g., between 145a1 and 431b1). That a revision supersedes another means that the claims in the superseded revision are replaced by the superseding revision. Only the owner of a revision is allowed to commit a superseding revision. Inclusion of one revision in another means that the included claims can be used as facts when applying rules.

For auditability, each claim in a revision is stored with evidence to make it possible to reconstruct how it was made. The inclusion relation between revisions allows us to include claims by other principals when applying rules. All claims from an included revision may be used as facts when applying rules.

\subsection{Monitoring with Revisions}


To integrate the revision model into our monitoring framework, we regard the local KB of each monitor as an incomplete revision, called \lstinline+staging+. When the monitors start, \lstinline+staging+ is empty. As before, a monitor can add claims to \lstinline+staging+ or facts can be derived by applying rules. 

A \lstinline+staging+ revision is transient, and needs to be \texttt{committed} to be stored persistently in the claim database. When committed, the claims from \lstinline+staging+ become a new revision~$R$ in the claim database. After a commit, the monitor starts with a fresh \lstinline+staging+ revision that will supersede~$R$ when it is committed. 

We allow the user to define how the fresh \lstinline+staging+ revision should be initialized from~$R$ by means of \emph{next-rules} in Cyberlog rulesheets. A next-rule has the form \lstinline+next a :- b1, ..., bn+ and is applied after the commit of a \lstinline+staging+ revision. The claims \lstinline+b1,...,bn+ are claims of the committed revision. This rule has the effect of adding the claim \lstinline+a+ to the new staging revision after a commit.

As an example, recall the example from Listing~\ref{lst:rulesheet} above. The following example for a Cyberlog next-rule specifies that a request should stay in the staging revision after a commit when it is still in process.
\begin{lstlisting}
 next request(Id, Data, TimeRequest) :- 
   request(Id, Data, TimeRequest),
   in_process(Id).
\end{lstlisting}

The monitors are configured to commit their revisions periodically, so that each monitor produces a linear chain of revisions, the last one of which contains the latest set of claims of the monitor.

To access the claims of other monitors, we use the feature that facts of other revisions can be included for reasoning by specifying the id of the revision to be included. The facts of the corresponding revision are then retrieved from the claim database and loaded in the corresponding KB. If the included revision is, at any time, superseded by another revision in claim database, then its claims and all consequences will be removed and be replaced with the claims of the \emph{new} revision.

\section{Implementation and Experimental Results}

To evaluate the monitoring approach in practice, we have implemented it with Kubernetes\footnote{https://www.kubernetes.io} and the Istio service mesh\footnote{https://www.istio.io}.

\textit{Claim database.}
Our implementation of the claim database is based on the tamper-proof log Trillian. Trillian implements an append-only storage and uses a Merkle-tree data structure for preventing tampering and for providing inclusion proofs, as  discussed in Section~\ref{sec:monitor-auditability}. It is used in large-scale applications, such as Certificate Transparency or sigStore\footnote{https://www.sigstore.dev}.

\textit{Security Monitors.} The security monitors implement a Cyberlog interpreter and the revision model. They are implemented in OCaml and use an efficient in-memory implementation of Datalog\footnote{https://github.com/c-cube/datalog} for performing reasoning. In the current prototype implementation the monitors are compiled to JavaScript using the ReScript compiler\footnote{https://rescript-lang.org/} and then executed using the NodeJS runtime. This makes it possible to use libraries not available in OCaml, e.g., for verifying Trillian inclusion proofs.

\begin{wrapfigure}{R}{0.3\textwidth}
    \centering
    \includegraphics[scale=0.4]{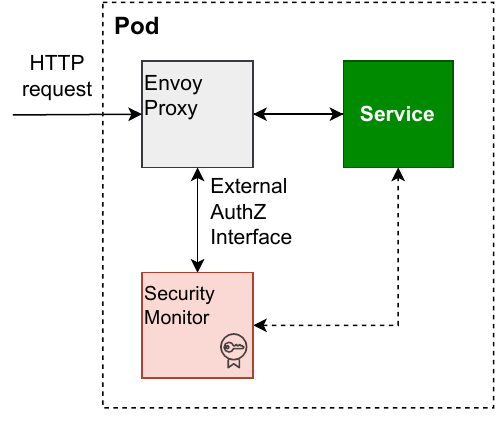}
    \caption{Security monitor integration using Istio's Envoy proxy}
    \label{fig:istio-integration}
\end{wrapfigure}
\textit{Cloud Integration.}
To apply the monitoring approach to a cloud system, the security monitors must be placed into the system to be able to observe events and communicate with the components under observation. Our implementation integrates transparently with a standard infrastructure built from Kubernetes and Istio. Kubernetes is a system for orchestrating containerized applications. Istio extends Kubernetes with a programmable software-defined network. Istio deploys with each application component a proxy container that is responsible for routing traffic, monitoring performance, enforcing authorization policies, and similar tasks. We deploy the security monitors together with the proxy containers with each application component. We use the \emph{external authorization service interface} of the proxy container to give the monitors access to the network traffic. This allows us to make network traffic available to the monitor and gives us the option to make authorization decisions in the monitors.
Fig.~\ref{fig:istio-integration} shows the integration of the security monitor inside a Kubernetes pod (a logical host machine).

HTTP requests to the application component go to the proxy, which forwards them to the security monitor. The monitor adds information about the request to its local KB. For example, a \texttt{POST} request generates the fact \lstinline+postRequest('/endpoint', body, timestamp)+ in the KB, where \lstinline+body+ is the message body. Recall that the monitor applies its Cyberlog rule continuously, so all consequences are also added to the KB.
 
We have applied the monitoring approach to the RTAPHM system. The rulesheets and configurations files are available\footnote{https://git.fortiss.org/sorokin/cyberlog-monitoring}.

\textit{Prototypical Evaluation.} To demonstrate the applicability of our approach, we have deployed the RTAPHM cloud system described in Section \ref{sec:use-case} in an on premise kubernetes cluster with k3s\footnote{https://k3s.io
}. The service components inside the cloud have been implemented by different partners with different technologies.
We initiated the scheduling process of a UAV through simulating requests by directly sending API requests to the SB. Outside the cluster we did deploy prototypically a DOM application, which has been connected to a UAV simulation (Gazebo). 

We have triggered several parallel booking processes and evaluated the \textit{processing delay} caused by forwarding API requests to security monitors. Further, we measured the amount of facts stored in each KB of a monitor. The minimal delay was 2ms (at SB), while the average delay was 29ms. The maximal introduced delay among all monitors was 113ms at MRM which is relatively high. This is not surprising, as in the rules of the MRM monitor mostly JSON arrays have to be processed and computation time is required to generate facts for each element of an array. Further, the minimal overall delay was 135ms, while the maximal was 157ms. The execution delays can further be reduced by parallelizing the fact reasoning in the security monitor. Considering the number of facts, the minimal KB size was 163 (at SB), while the maximal was 259 (at CA). 
After one service request the KB size increased maximally by less than 184 facts (maximal fact size is 15KB). The processing of the security monitor further required 1mCPU which is negligible. 

\section{Related Work}
We give a brief overview of related work for the runtime monitoring of distributed and web-based applications and describe some work that considers the auditability of system events. 

\subsection{Runtime monitoring}



Runtime monitoring of distributed applications has been exhaustively studied. Many approaches are based on generating automata from temporal logics specifications or using stream processing \cite{rtConversationChechik,cotroneo2020runtime,BratanisMonitoring10,DeckerMedMon14,sen_decentral,BratanisMonitoring10}. For instance, \cite{sen_decentral} introduces a distributed specification language to generate distributed monitors. 
Bauer et al.~\cite{bauer_decent} address the distributed synthesis of monitors from LTL specifications based on system execution traces.
In \cite{rtConversationChechik}, the authors present an approach to generate non-deterministic finite state automata for monitoring global liveness and safety properties of web services. 
Automata-based approaches cannot be easily extended to support the monitoring of parallel processes, as well monitoring properties that depend on the data in the exchanged messages.

 Cotroneo et al. have proposed an approach \cite{cotroneo2020runtime} for monitoring a cloud system using stream processing. API requests are first logged in a fault-free system execution to learn the correct behaviour of the system in form of rules to observe later whether the system obeys the rules.
 Also in this approach, it is not possible to specify properties that consider data in the API requests. Also, monitoring results are not provided back to the distributed components, to control the execution of the component based on the monitoring results.

Our monitoring approach is related to the authorization framework Open Policy Agent
(OPA)\footnote{https://www.openpolicyagent.org}, whose idea is to decouple the enforcement of authorization policies from the application code. Policies are specified in a declarative language called Rego which is similar to Cyberlog. While OPA has an interface to integrate external data when applying rules, processed and reasoned data is not auditable and no attestions are supported. Since OPA is only an authorization framework, no monitoring of workflow events is possible. 


\subsection{Auditability of system events}

Google's Certificate Transparency (CT) framework\footnote{https://certificate.transparency.dev} is a prominent example for making decisions based on auditable data: the goal is to identify falsely or maliciously issued certificates, by recording all HTTPS certificates issued
by publicly trusted Certificate Authorities in a tamper-proof log called Trillian\footnote{http://transparency.dev/\#trillian}. Trillian is a scalable alternative to distributed, tamper-proof data storages, such as the blockchain. The concept has been adopted by several projects, whether for supply chain security in sigStore, identification of malicious firmware with Binary Transparency~\cite{trillian_ft}, for the detection of malicious software updates from package managers~\cite{Hof2017} or the verifiable auditing of (confidential) data accesses~\cite{VamsHicks2018}.

An extension of standard SIEM approaches is provided by syslog-ng%
\footnote{https://www.syslog-ng.com/} which assures the secure transmission and tamper-proof storage of logged data to meet compliance requirements. Nevertheless, it is not possible to monitor requirements which depend on this data, so that components can incorporate the monitoring results in their decisions.

Another work \cite{Prybila2020RuntimeVF} has
applied blockchain for the runtime monitoring of business processes. Recorded events are auditable, but the transaction time takes more than 7 minutes. This approach is therefore only applicable in a use case with long-running tasks, such as logistics or manufacturing. Further, a blockchain deployment is complex and requires a considerable amount of resources, which is in contrast to our approach when using Trillian.

In \cite{peerreview} the authors have presented an approach making distributed system auditable using tamper-proof storage focusing on the detection of byzantine faults. Similiar to our approach, monitored system components have to cryptographically sign messages, and events are recorded in the tamper-proof log. However, the approach does not allow the monitoring of workflows based on auditable claims as well as a flexible specification of monitored properties.





\section{Conclusion} 
We have proposed a new monitoring approach for monitoring safety and data-related properties in cloud applications. To satisfy the distributed nature of deployed systems in a cloud, we have introduced Cyberlog for the specification of monitored properties as an extension of Datalog by using attestations. Monitoring is then performed by the execution of distributed Cyberlog programs. The approach is non-intrusive because only communication between services is processed.
Further, we presented an approach for the revision of claims inspired by file revision systems such as git, to handle the problem of increasing storage of runtime events which comes by using a Datalog-based language. To allow for the documentation of the processed information to satisfy regulatory requirements our approach integrates a scalable tamper-proof ledger to store all processed and reasoned monitoring claims. 
We prototypically evaluated our monitoring approach on a cloud system for a safety-critical use case. 
Nevertheless, the specification of Cyberlog programs requires some effort. We are working on improving the syntax for the specification of properties and investigating approaches for synthesizing Cyberlog-programs from API specifications and use case information.
We assume that our monitoring approach and further research can increase the security of cloud systems and promote the deployment of applications for mission-critical systems in the cloud.

\section*{Acknowledgment}
This work was supported by the Federal Ministry for Economic Affairs
and Climate Action in the RTAPHM project, grant 20X1736M (Lufo V-3).

\bibliographystyle{splncs04} 
\bibliography{paper.bib}

\begin{thebibliography}{10}
\providecommand{\url}[1]{\texttt{#1}}
\providecommand{\urlprefix}{URL }
\providecommand{\doi}[1]{https://doi.org/#1}

\bibitem{ahmad_secure_2019}
Ahmad, A., Saad, M., Mohaisen, A.: Secure and transparent audit logs with
  blockaudit. Journal of Network and Computer Applications  \textbf{145},
  102406 (2019)

\bibitem{DeclarativeCloud}
Alvaro, P., Condie, T., Conway, N., Elmeleegy, K., Hellerstein, J.M., Sears,
  R.: Boom analytics: Exploring data-centric, declarative programming for the
  cloud. In: European Conference on Computer Systems (EuroSys'10). p.
  223–236. ACM (2010)

\bibitem{Dedalus}
Alvaro, P., Marczak, W., Conway, N., Hellerstein, J., Maier, D., Sears, R.:
  Dedalus: Datalog in time and space. In: Datalog Reloaded. pp. 262--281.
  Springer (2011)

\bibitem{BachaneCloudSIEM16}
Bachane, I., Adsi, Y.I.K., Adsi, H.C.: Real time monitoring of security events
  for forensic purposes in cloud environments using {SIEM}. In: Systems of
  Collaboration (SysCo 2016). pp.~1--3 (2016)

\bibitem{bauer_decent}
Bauer, A., Falcone, Y.: Decentralised {LTL} monitoring. Form.~Meth.~Syst.~Des.
  \textbf{48} (2016)

\bibitem{BratanisMonitoring10}
Bratanis, K., Dranidis, D., Simons, A.J.H.: An extensible architecture for
  run-time monitoring of conversational web services. In: Monitoring,
  Adaptation and Beyond (MONA 2010). p. 9–16. ACM (2010)

\bibitem{CassarRV17}
Cassar, I., Francalanza, A., Aceto, L., Ing{\'{o} }lfsd{\'{o}}ttir, A.: A
  survey of runtime monitoring instrumentation techniques. EPTCS  \textbf{254},
   15--28 (2017)

\bibitem{datalogCeri89}
Ceri, S., Gottlob, G., Tanca, L.: What you always wanted to know about datalog
  (and never dared to ask). IEEE Trans. Knowl. Data Eng.  \textbf{1},  146--166
  (1989)

\bibitem{ChuDSN07}
Chu, D., Popa, L., Tavakoli, A., Hellerstein, J.M., Levis, P., Shenker, S.,
  Stoica, I.: The design and implementation of a declarative sensor network
  system. In: Embedded Networked Sensor Systems (SenSys 2007). p. 175–188.
  ACM (2007)

\bibitem{cotroneo2020runtime}
Cotroneo, D., Simone, L.D., Liguori, P., Natella, R., Scibelli, A.: Towards
  runtime verification via event stream processing in cloud computing
  infrastructures (2020)

\bibitem{DeckerMedMon14}
Decker, N., Kühn, F., Thoma, D.: Runtime verification of web services for
  interconnected medical devices. In: ISSRE 2014. pp. 235--244 (2014)

\bibitem{rtConversationChechik}
Gan, Y., Chechik, M., Nejati, S., Bennett, J., O'Farrell, B., Waterhouse, J.:
  Runtime monitoring of web service conversations. In: Computer Science and
  Software Engineering (CASCON 2017). p. 2–17. {IBM} / {ACM}, USA (2017)

\bibitem{trillian_ft}
Google: Firmware transparency (2021),
  \url{https://github.com/google/trillian-examples/tree/master/binary\_transparency/firmware}

\bibitem{peerreview}
Haeberlen, A., Kouznetsov, P., Druschel, P.: Peerreview: Practical
  accountability for distributed systems. SIGOPS Oper. Syst. Rev.
  \textbf{41}(6),  175–188 (Oct 2007)

\bibitem{miller2011harper}
Harper, A., VanDyke, S., Blask, C., Harris, S., Miller, D.: Security
  information and event management (siem) implementation  (2010)

\bibitem{VamsHicks2018}
Hicks, A., Mavroudis, V., Al-Bassam, M., Meiklejohn, S., Murdoch, S.J.: {VAMS:}
  verifiable auditing of access to confidential data. CoRR
  \textbf{abs/1805.04772} (2018)

\bibitem{Hof2017}
Hof, B., Carle, G.: Software distribution transparency and auditability. CoRR
  \textbf{abs/1711.07278} (2017)

\bibitem{LevesonSTPA04}
Leveson, N.: A new accident model for engineering safer systems. Safety Science
   \textbf{42}(4),  237--270 (2004)

\bibitem{Prybila2020RuntimeVF}
Prybila, C., Schulte, S., Hochreiner, C., Weber, I.: Runtime verification for
  business processes utilizing the bitcoin blockchain. Future Gener. Comput.
  Syst.  \textbf{107},  816--831 (2020)

\bibitem{sen_decentral}
Sen, K., Vardhan, A., Agha, G., Rosu, G.: Efficient decentralized monitoring of
  safety in distributed systems. In: ICSE 2004. pp. 418--427 (2004)

\bibitem{van_hoye_logging_2019}
Van~Hoye, L., Maenhaut, P.J., Wauters, T., Volckaert, B., De~Turck, F.: Logging
  mechanism for cross-organizational collaborations using {Hyperledger}
  {Fabric}. In: {Int.} {Conf.} on {Blockchain} and {Cryptocurrency} ({ICBC}
  2019). pp. 352--359. IEEE (2019)

\end{thebibliography}

\end{document}